\documentclass[fleqn,twoside]{article}
\usepackage{espcrc2}

\usepackage{graphicx}
\usepackage[figuresright]{rotating}

\newcommand{\AmS}{{\protect\the\textfont2
  A\kern-.1667em\lower.5ex\hbox{M}\kern-.125emS}}

\def\cL{{\cal L}} \def\cM{{\cal M}}

\def\GeV{\mathrm{GeV}}

\newcommand{\ga}{\gamma}

\newcommand{\beq}{\begin{equation}}
\newcommand{\eeq}{\end{equation}}
\newcommand{\ba}{\begin{array}}
\newcommand{\ea}{\end{array}}
\newcommand{\bea}{\begin{eqnarray}}
\newcommand{\eea}{\end{eqnarray}}

\renewcommand{\Im}{\mathrm{Im}}
\renewcommand{\Re}{\mathrm{Re}}
\newcommand{\phit}{\varphi_T}
\newcommand{\phis}{\varphi_S}

\newcommand{\nuu}{$\nu$ }
\newcommand{\nuus}{$\nu$s }

\title{LFV in Models with A4 Flavour Symmetry}

\author{Luca Merlo\address{Dipartimento di Fisica `G.~Galilei', Universit\`a di Padova\\
INFN, Sezione di Padova, Via Marzolo~8, I-35131 Padua, Italy\\
{\rm E-mail: merlo@pd.infn.it}}}

\begin{document}

\begin{abstract}
The approximated tri-bimaximal mixing (TBM) observed in the $\nu$ oscillations is a particular feature of a class of models characterized by the spontaneously broken horizontal flavour symmetry $A_4$. In this paper, it is presented an analysis on the predictions of these models for relevant low-energy observables.
In an effective operator approach, these effects are dominated by dim-6 operators, suppressed by the scale $M$ of new physics (NP).
Indications for $M$ and an upper bound on $\theta_{13}$ of a few percent are found.
\vspace{1pc}
\end{abstract}

\maketitle


\section{INTRODUCTION}
The Standard Model (SM) fails in describing nature and its behaviour: \nuus are the most outstanding proof of this defeat, in fact the solar and atmospheric anomalies find the simplest solution in the \nuu oscillations, possible only if \nuus are massive particles. However SM represents the starting point of almost all beyond the Standard Model (BSM) theories, due to its predictivity in many physical sectors.
It is then fundamental to understand what is the theory which can embed SM and describe correctly \nuus at the same time.\\
In order to overcome our ignorance of the correct BSM theory, a very appealing description can be done by a low-energy effective approach, where the Lagrangian is a sum of the usual SM one plus non-renormalizable terms, suppressed by the correct power of the cut-off scale $\Lambda$:
\begin{equation}
\cL=\cL_{SM}+\frac{\cL_5}{\Lambda}+\frac{\cL_6}{\Lambda^2}+\ldots
\end{equation}
Here, $\cL_6$ represents more than 80 independent dim-6 operators, while $\cL_5$ the unique, up to flavour combination, dim-5 operator and it is responsible for the \nuu masses:
\begin{equation}
\frac{\cL_5}{\Lambda}=\frac{1}{\Lambda}({\tilde H}^\dagger l)^T Y  ({\tilde H}^\dagger l)+h.c.=\frac{v^2}{2\Lambda}\nu^T Y \nu+h.c.
\end{equation}
where $Y$ is a 3 $\times$ 3 complex symmetric matrix. From a simple dimensional estimation, \mbox{$\Lambda\simeq10^{15}\GeV$}, therefore we can say that \nuus provide a window in the GUT physics.\\
The simplest scenario we can imagine consists in the presence of two energy scales: a first very large, $M_{GUT}$, where
GUTs and flavour symmetries find their natural settlement; a second very low, the e.w. scale, at which particle masses and mixing angles have the measured values.
It is very difficult to get informations about the theory at $M_{GUT}$ from the low-energy observables, therefore the search for new observables becomes underlying. A possibility is to introduce an intermediate energy scale, $M$, at about $1-10$ TeV: this corresponds to the presence of some kind of NP, which we do not specify, at this scale. Other indications, which enforce this choice, come from
other sectors of Physics, like the anomalous $(g-2)_\mu$, Dark Matter, the running of the gauge coupling constants and the hierarchy problem, which would benefit by the introduction of this scale $M$.\\
In section \ref{ModelsA4}, we illustrate the construction of a model based on the discrete non-Abelian $A_4$ group, whose relevant feature is to predict the TBM\cite{Chen}, that is an excellent approximation of the experimental data\cite{Palazzo}: at $2\sigma$ errors
\beq
\ba{ll}
\sin^2\theta_{13}<0.036 & \sin^2\theta_{13}^{TBM}=0\\
\sin^2\theta_{23}=0.466^{+0.136}_{-0.100} & \sin^2\theta_{23}^{TBM}=1/2\\
\sin^2\theta_{12}=0.312^{+0.040}_{-0.034} & \sin^2\theta_{12}^{TBM}=1/3\;.
\ea
\eeq
In section \ref{Predictions}, we analyze the predictions of the model for a set of relevant low-energy observables, as lepton flavour violating (LFV) transitions, leptonic magnetic dipole moments (MDM) and electric dipole moments (EDM), in two separate cases, a supersymmetric and a general one. The results are summarized in section \ref{Conclusion}.

\section{MODELS WITH $A_4$ FLAVOUR SYMMETRY}
\label{ModelsA4}

Several models have been proposed to produce the TBM scheme (a complete list of papers in \cite{FHLM_LFV}) and among the most economic and simplest ones are those based on the discrete group $A_4$\cite{AF_ED,AFL_Orbifold}. $A_4$ is the group of even permutations of four objects.
It is generated by two elements
$S$ and $T$ obeying the relations $S^2=(ST)^3=T^3=1$.
The group $A_4$ has two obvious subgroups: 
$G_S$ isomorphic to $Z_2$, generated by $S$, and 
$G_T$ isomorphic to $Z_3$, generated by $T$. They are the relevant low-energy symmetries
of the \nuu and the charged-lepton sectors at leading order, respectively. The TBM is then a direct consequence of this
special symmetry breaking pattern, which is achieved via the vacuum misalignment of triplet scalar fields, called flavons.\\
Concerning the flavour group $G_f$, following \cite{AF_ED}, we choose
\beq
G_f=A_4\times Z_3\times U(1)_{FN}\;.
\eeq
The three factors in $G_f$ play different roles. The spontaneous breaking of $A_4$ is directly responsible for the TBM, while the $Z_3$ factor is a discrete version of the total lepton number. Finally, $U(1)_{FN}$ is
responsible for the hierarchy among the charged fermion masses.
The flavour symmetry breaking sector of the model includes the scalar fields $\varphi_T$, $\varphi_S$, $\xi$ and $\theta$: \mbox{$\langle\phit\rangle\propto(u,0,0)$}, which breaks $A_4$ down to $G_T$, \mbox{$\langle\phis\rangle\propto(u,u,u)$},which breaks $A_4$ down to $G_S$, $\langle\xi\rangle\propto u$ and $\langle\theta\rangle=t$, where $u$ and $t$ are the small symmetry breaking parameters of the theory, respectively $0.001<|u|<0.05$
and $|t|\simeq0.05$. This VEV alignment comes from the minimization of the scalar potential (see \cite{AF_ED}).\\
Therefore with a particular assignment of the quantum numbers to the SM particles and the flavons and the well defined vacuum alignment just described, the model predicts, at the leading order, a diagonal charged lepton mass matrix and a \nuu mass matrix which is exactly diagonalized by the TBM matrix. The sub-leading terms introduce corrections to the mass matrices of relative order $|u|$
providing a non-zero value of $\theta_{13}$ and a non-maximal value for $\theta_{23}$. Such a framework can also be extended to the quark sector\cite{FHLM_Tprime}.


\section{PREDICTIONS FOR LOW-ENERGY OBSERVABLES}
\label{Predictions}

In an effective field theory approach, we first integrate out the d.o.f. related to $M_{GUT}$, giving rise to the flavour structure of ${\cal L}_{eff}$, and then the d.o.f related to $M$, therefore the dominant physical effects of the NP at low energies
can be described by dim-6 operators, suppressed by $M^2$ and not by $1/\Lambda^2$, opening the possibility that the related effects might be seen in the future.
These observables are the MDMs, the EDMs and the LFV transitions $\mu\to e \gamma$, $\tau\to\mu\gamma$ and $\tau\to e \gamma$.
In the lepton sector, the leading terms of the relevant effective Lagrangian are \cite{deGouvea}:
\begin{equation}
\cL_{eff}=\cL_{SM}+\delta \cL(m_\nu)+i\frac{e}{M^2} {e^c}^T H^\dagger \sigma F \cM\ell+\ldots
\end{equation}
where $e$ is the electric charge, $e^c$ the set of SU(2) lepton singlets,
$F_{\mu\nu}$ is the electromagnetic field strength and $\cM\equiv\cM\left(\langle\phi\rangle\right)$ is a complex $3\times 3$ matrix $\cM$, with indices in the flavour space.
The effective Lagrangian ${\cal L}_{eff}$ is invariant under $G_f$, once
we treat the symmetry breaking parameters as spurions. As a result, the same symmetry breaking parameters that control lepton masses and mixing angles also control
the flavour pattern of the other operators in $\cL_{eff}$ and we get that $\cM$ has the same flavour structure of the charged lepton Yukawa matrix. In the basis of canonical kinetic terms and diagonal charged leptons (we denote by a hat the relevant matrices in this basis), $\Re(\hat\cM_{ii})$ are proportional to MDMs, $\Im(\hat\cM_{ii})$ to EDMs and $\hat\cM_{ij}$ describe the LFV transitions.
We can derive a bound on $M$, by considering the existing bounds on MDMs and EDMs. The strongest constraint, $M>80$ TeV, comes from the $d_e$: in order to lower this value in the range we have previously indicated, we need to invoke a cancellation in $\Im[\cM]$, which could be accidental or due to some kind of CP-conservation. A very interesting indication for LHC comes from $\delta a_\mu$, $M\approx2.7$ MeV. Considering now the LFV transitions, 
$BR(\mu\to e\gamma)\simeq BR(\tau\to\mu\ga)\simeq BR(\tau\to e\ga)$: this represents a distinctive feature of the model. Therefore, imposing the existing bounds on $\mu\to e\gamma$, we can say that the $\tau$ decays are below the expected future sensitivity. Moreover we get $M>10-70$ TeV, depending on $|u|$. We finally conclude that the bounds we have found, are above the region of interest for $(g-2)_\mu$ and for LHC.

\subsection{Supersymmetric case}
The previous results are obtained without specifying the type of NP: if now we consider the presence of low-scale SUSY, we find new and interesting outcomes. The off-diagonal entries of $\cM$ come from two sources: the next-to-leading corrections to $\langle\phit\rangle$ and the double flavon insertion of the type $\xi^\dagger\phis$.  These contributions are in general of the same size, but if the theory is supersymmetric, with a softly broken SUSY, the first is dominant \cite{FHLM_LFV}.
The result is an additional suppression on the elements below the diagonal of $\cM$: the constraints from $d_e$ and $\delta a_\mu$ remain the same, but we find $BR(\mu\to e\gamma)\simeq BR(\tau\to\mu\ga)>BR(\tau\to e\ga)$ and that the corresponding bound on M is softened, $M>1-14$ TeV, depending on $|u|$. Now we can conclude that there is a range for $|u|$, in which the model explain the discrepancy in $(g-2)_\mu$ and a probably positive signal of $\mu\to e\ga$ at MEG.\\
We can eliminate the dependence on the unknown scale $M$:
\begin{equation}
BR_{\mu\to e\gamma}\!\propto\!\left(\delta a_\mu\right)^2\!\!\left[\vert \tilde{w}^{(1)}_{\mu e}\vert^2 \vert u\vert^4+\frac{m_e^2}{m_\mu^2} \vert \tilde{w}^{(2)}_{\mu e}\vert^2\vert u\vert^2\right]
\label{muegamma}
\end{equation}
We plot $BR(\mu\to e\gamma)$ versus $|u|$ in fig. 1, where $\delta a_\mu=30.2(8.8)\times10^{-10}$ and $\tilde{w}^{(1,2)}_{\mu e}$ are kept fixed to 1 (darker region) or are random numbers with $0<|\tilde{w}^{(1,2)}_{\mu e}|<2$ (lighter region). We cannot conclude a sharp limit on $|u|$, but we see that the present limit on $\mu\to e \gamma$ disfavors values of $|u|$ larger than few percents. The same we can say for $\theta_{13}$, since it is comparable to $|u|$.
\begin{figure}[htb]
\vspace{-0,3cm}
\includegraphics[scale=0.25]{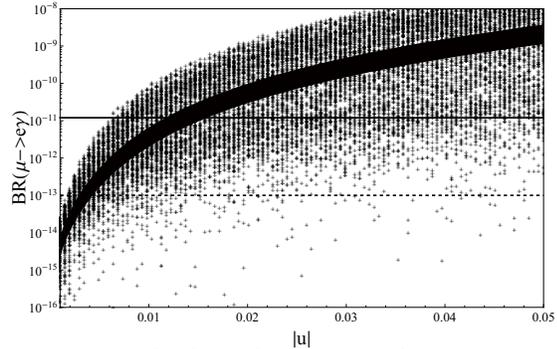}
\vspace{-1,8cm}
\caption{The branching ratio of $\mu\to e \gamma$ as a function of $|u|$, eq. \ref{muegamma}.}
\vspace{-0,7cm}
\end{figure}


\section{CONCLUSION}
\label{Conclusion}

The lepton mixing matrix is well approximated by the TBM, which is easily recovered by flavour models based on the discrete group $A_4$: among the others, $A_4\times Z_3\times U(1)_{FN}$ represents a sort of minimal choice. We have studied the predictions of this model for a set of observables and we can conclude that the supersymmetric version suggests the presence of NP at about a few TeV, which explains the discrepancy in $(g-2)_\mu$ and a probably positive signal for $\mu\to e\ga$ at MEG and indicates an upper bound for $\theta_{13}$ of few percents.

\section*{Acknowledgements}
I thank the organizers of NOW 2008
for giving me the opportunity to present my talk and for the kind hospitality in the Salento peninsula.
Alike I thank Ferruccio Feruglio, Claudia Hagedorn and Yin Lin for the advantageous collaboration.

\end{document}